# Plasma treatment advantages for textiles

**Amelia Sparavigna**
Dipartimento di Fisica, Politecnico di Torino
Corso Duca Abruzzi 24, Torino, Italy

**Abstract**
The textile industry is searching for innovative production techniques to improve the product quality, as well as society requires new finishing techniques working in environmental respect. Plasma surface treatments show distinct advantages, because they are able to modify the surface properties of inert materials, sometimes with environment friendly devices. For fabrics, cold plasma treatments require the development of reliable and large systems. Such systems are now existing and the use of plasma physics in industrial problems is rapidly increasing. On textile surfaces, three main effects can be obtained depending on the treatment conditions: the cleaning effect, the increase of microroughness (anti-pilling finishing of wool) and the production of radicals to obtain hydrophilic surfaces. Plasma polymerisation, that is the deposition of solid polymeric materials with desired properties on textile substrates, is under development. The advantage of such plasma treatments is that the modification turns out to be restricted in the uppermost layers of the substrate, thus not affecting the overall desirable bulk properties. Here, we present the research results on the use of plasma physics in textile properties modification. Treatments on natural, wool and cotton, and on synthetic polymers to improve wetting are shown. Hydrophilic-hydrophobic treatments, dirt-repellent coatings are presented. Low-pressure and atmospheric-pressure glow discharge systems are also discussed.

**Introduction**
The textile industry is concerned with the following interrelated issues: development and production of raw materials, converting them into finished products, meeting customer expectations with human health and environmental safety, and making a profit. Many of these issues parallel the general priorities of society, that is, sustainable production systems in environment safety, economies of rural and industrial communities, and consumer interests.

Finding new products, alternative raw materials, and processing technologies more "environmentally friendly", obtained with new natural processes and conditions, is central for further developments of textile industry [1]. Moreover, the production needs innovative recycling technologies for wastes, because the amount of post-consumer textiles annually generated turns out to be considerable for the increasing environmental concerns.
Since their introduction, synthetic fibers have a significant impact on the quality of life.

Nevertheless the demand for natural fibers continues to increase, for their many outstanding properties including aesthetics, comfort, and biodegradability. Besides research efforts on traditional natural materials that is cotton, wool and silk, many researches are also focused today on exploring fibers from alternative crops and agricultural by-products, which are often underutilized. The abundance of recyclable natural fibers and agricultural residues drive researchers to develop alternative structural composites for textile applications [2-4].

Using natural or synthetic fibers, with old or new agricultural materials, the textile industry must go towards sustainable technologies, developing environmentally safer methods of processing and finishing fabrics. One way, now under study, is the processing with biological systems, rather than conventional chemistries. This new frontier in biotechnology for textile processing includes cross-linking in polymers to impart easy-care properties, surface modification to enhance absorption/dyeing and aesthetic properties, and enzyme systems for scouring, antipilling and for imparting anti-felting properties to wool [5]. Moreover to reduce energy consumption, chemicals, and time waste in textile processing, the use of multi-purpose dyes and finishing agents (for instance, dual purpose dye-insect control agents) are under development.

Optimization of bulk and surface properties of materials can represent a promising approach for meeting technical and economical requirements. Because of costs related to study and production of new fibers, polymer researchers now focus on modifying existing fibers to impart the desired aesthetic or functional properties. Conventional fibre modification methods include various thermal, mechanical, and chemical treatments.

Another important method to modify the fibre, to increase the uptake of dyes and finishes or to impart unique functionality, is performed through cold plasma [6-8]. The reactive species of plasma, resulting from ionization, fragmentation, and excitation processes, are high enough to dissociate a wide variety of chemical bonds, resulting in a significant number of simultaneous recombination mechanisms. Well known in semiconductor physics, plasma opens up new possibilities for polymer industrial applications where the specific advantages of producing pore-free, uniform thin films of superior physical, chemical, electrical and mechanical properties have been required.

For instance, main advantages of a plasma polymerization methods are: 1) applicability to almost all organic, organo-metallic and hetero-atomic organic compounds, 2) modification of surface properties without altering the bulk characteristics, 3) low quantities needed of monomeric compounds making it non energy intensive, and 4) wide applicability to most organic and inorganic structures. Exciting plasma applications include 1) cold plasma discharge synthesis of new polymeric structures, 2) plasma induced polymerization processes, 3) surface grafting of polymers, and 4) surface modification of polymers. Characteristics that can be improved include wettability, flame resistance, adhesive bonding, printability, electromagnetic radiation reflection, surface hardness, hydrophilic-hydrophobic tendency, dirt-repellent and antistatic properties.

Plasma treatments can answer the demand of textile industry. Besides the base function of dressing people, textiles contribute to human health and safety, protecting from exposure to dangerous environments. Important requirements for protective apparel are barrier effectiveness and comfort for the wearer. The effectiveness of protective apparel worn for instance by patients and health care workers continues to be a major concern of textile industry, because of the risks of infection when the protective apparel fails. An effective protection in medical applications demands higher-quality and reliable adhesive bonds, and the plasma treatment, as we will discuss in this paper, is providing enhanced adhesive bond strength and permanency for medical devices and disposables.

Researches on modifying the surface characteristics of fibers using plasma technology are performed by many scientific groups. A special plasma application, in fact rather relevant for textile industries, is that developed by Sarmadi and his group, at the Plasma Research Institute of the Wisconsin University. They are evaluating multifunctional reactive dyes and plasma treatments for imparting insect control. Wool and specialty hair fibers are readily attacked by clothes moth and larvae [9]. With bans on some insecticide materials [10], alternative control methods are needed. Using plasma is a novel approach that reduces air, water and land pollution in comparison to conventional methods of wet chemistry.

Plasma surface treatment used to modifying the functional properties of fibers possesses advantages in comparison with traditional techniques. Plasma includes less water usage and energy consumption, with a very small fibre damage, then making plasma process very attractive. It will be used to enhance the quality of textile products in fabric preparation and in dyeing and finishing methods. In this paper, recent developments in the plasma treatment on textile surfaces are presented. In the first part a brief discussion on cold plasma is performed then followed by some case studies. The last part contains a description of plasma industrial devices, in particular of atmospheric pressure devices.

**Plasma processes.**

Faraday proposed to classify the matter in four states: solid, liquid, gas and radiant. Researches on the last form of matter started with the studies of Heinrich Geissler (1814-1879): the new discovered phenomena, different from anything previously observed, persuaded the scientists that they were facing with matter in a different state. Crookes took again the term "radiant matter" coined by Faraday to connect the radiant matter with residual molecules of gas in a low-pressure tube. Sufficient additional energy, supplied to gases by an electric field, creates plasma. For the treatment of fabrics, cold plasma is used, where the ambient treatment atmosphere is near room temperature. It can be produced in the glow discharge in a vacuum process or in more recent atmospheric pressure plasma devices.

Plasma is partially ionized gas, composed of highly excited atomic, molecular, ionic and radical species with free electrons and photons. In cold plasma, although the electron temperature can be much more high, the bulk temperature is essentially the ambient one. Plasma can be obtained between electrodes in high frequency devices (typically 40 kHz or 13.56 MHz) or with microwave generators (2.45 GHz).

Let us start our discussion on the plasma processes performed to modify fibers and polymer surfaces (wool, silk, cotton are natural polymers). Plasma processes can be conveniently classified into four overall processes: *cleaning, activation, grafting* and *deposition.*

In a *cleaning process*, inert (Ar, He) and oxygen plasmas are used. The plasma-cleaning process removes, via ablation, organic contaminates such as oils and other production releases on the surface of most industrial materials. These surface contaminants as polymers, undergo abstraction of hydrogen with free radical formation and repetitive chain scissions, under the influence of ions, free radicals and electrons of the plasma, until molecular weight is sufficiently low to boil away in the vacuum (see fig.1).

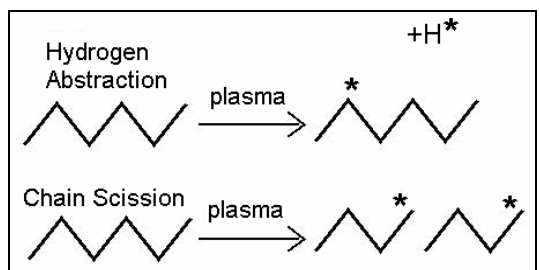

Fig.1: Free radicals formation by means of plasma action. Plasma can abstract hydrogen from the polymeric chain or can split chains.

*Activation plasma processes* happen when a surface is treated with a gas, such as oxygen, ammonia or nitrous oxide and others, that does not contain carbon. The primary result is the incorporation of different moieties of the process gas onto the surface of the material under treatment. Let us consider the surface of polyethylene, which normally consists solely of carbon and hydrogen: with a plasma treatment, the surface may be activated, anchoring on it functional groups such as hydroxyl, carbonyl, peroxyl, carboxylic, amino and amines (Fig. 2). Hydrogen abstraction produces free radicals in the plasma gases and functional groups on the polymeric chain.

Almost any fibre or polymeric surface may be modified to provide chemical functionality to specific adhesives or coatings, significantly enhancing the adhesion characteristics and permanency. For instance, polymers activated in such a manner provide greatly enhanced adhesive strength and permanency, and this is a great improvement in the production of technical fabrics.

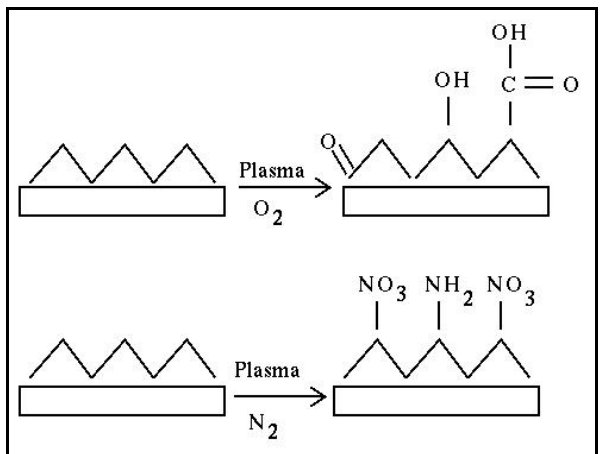

Fig.2: An example of surface activation by substituting hydrogen in a polymeric chain with other groups such as O, OH, COOH, $NO_3$, $NH_2$, etc.

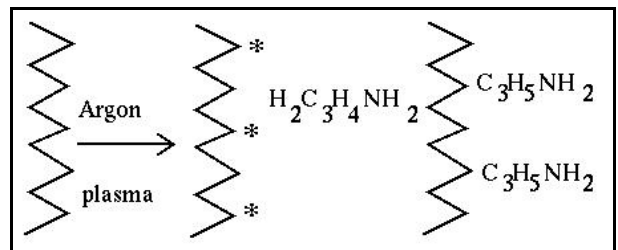

Fig.3: Grafting of a monomer on the surface: argon plasma produces radicals on the chain and monomers are grafted on the surface.

Plasma can also produce a material *deposition*: when a more complex molecule is employed as the process gas, a process known as plasma-enhanced chemical-vapour deposition (PECVD) may result. For instance, when methane or carbon tetrafluoride is employed, the gas undergoes fragmentations in the plasma, reacting with itself to combine into a polymer. Selecting the process conditions, pinhole-free chemically unique films, may be deposited onto surfaces of materials within the plasma reactor. PECVD coatings alter in a permanent way the surface properties of the material onto which these have been deposited.

In *grafting*, an inert gas such as argon is employed as process gas, many free radicals shall be created on the material surface. If a monomer capable of reacting with the free radical is introduced into the chamber, the monomer shall become grafted. This is a procedure for low-pressure plasma treatment but grafting can be obtained with atmospheric plasma processing too. Typical monomers are acrylic acid, allyl amine and allyl alcohol (Fig. 3).

By means of the plasma processes some properties of the surface can be changed to obtain several applications. First of all, surface wetting changes. Polymeric surface are usually not wettable and adhesion is poor. After plasma treatment the surface energy increases and wettability and adhesion enhancements are produced.

Whether it is an adhesive-bonded coating or a painted or printed decoration and markings, plasma surface treatment can provide significant improvement. The strength of an adhesive coating is enhanced by several factors: removal of contaminants, an increase in the surface energy of the substrate and consequently in wetting, and the formation of chemically reactive sites for the adhesive or ink that covalently bond. Covalent bonds provide stability because bonds are scarcely affected by aging, ultraviolet-light exposure or other environmental effects. Plasma can provide improvements when gas employed as well as plasma parameters are optimized for each material and adhesive class [11-13].

Several materials have been plasma treated to increase bond strength till a bond permanency. In most instances, the loci of failure shifts with plasma treatment from adhesive-bond failure to cohesive failure, either within the adhesive layer or the substrate being bonded. Adhesion strength improvements are usually studied for commercial epoxy adhesives.

Fluoropolymers are typically considered as materials to which quality adhesive bonds are impossible. Hydrogen plasma, often in conjunction with a co-process gas, has been found to be particularly effective. Plasma processes employing hydrogen cause dehydrohalogenation along the fluoropolymer backbone to which the co-process gas can subsequently covalently attach [14].

Depending on the gas employed, plasma treatment may render the surface very hydrophilic, oleophobic or hydrophobic. The hydrophilic feature can be controlled very well. Oleo- and hydrophobic features are readily achieved in plasmas containing fluorine.

If a fluoro-alkane such as tetrafluoromethane or hexafluoroethane is utilized as process gas, fluorine will be substituted for abstracted hydrogen on the surface of the substrate, reducing its surface energy. Plasma-induced grafting offers another method of providing specific reactive sites to normally inert polymers. When an unsaturated monomer such as allyl alcohol or allyl amine is introduced into the reaction low-pressure chamber after the plasma is extinguished and prior chamber ventilation, it will add to the free radical, yielding a grafted polymer. If plasma in vacuum chamber is used, the treatment must be considered as a batch treatment. On-line atmospheric systems can be used too.

Plasma systems of all sizes and degrees of automation are now available, from manually loaded batch systems as small as a microwave oven to fully automated systems treating automotive instrument panels (a section will be devoted to the discussion of industrial devices).

To summarize this section we can tell that cold gas plasma processes offer an efficient and reliable means to alter surface properties of all materials without affecting the bulk properties of the treated material. Reengineering the surfaces by introducing functional groups in a controlled and reproducible manner, greatly enhances adhesive permanency and reliability. The nature of cold gas plasma surface modification leads itself on precise control and process repeatability. In the vast majority of applications, plasma surface treatment employs innocuous gases, enabling the manufacturing engineer to avoid corrosive chemicals and solvents.

**Textile applications.**

The research in the field of plasma applications for textile treatments is very wide and we will discuss some recent developments. Due to the great amount of literature results only a few exemplary applications with related plasma gases are summarized in the following.

*1) Enhance mechanical properties.* Softening of cotton and other cellulose-based polymers, with a treatment by oxygen plasma. Reduced felting of wool with treatment by oxygen plasma. Top resistance in wool, cotton, silk fabrics with the following treatment: dipping in DMSO and subsequently $N_2$-plasma.

*2) Electrical Properties.* Antistatic finish of rayon, with chloromethyl dimethylsilane in plasma.

*3) Wetting.* Improvement of surface wetting in synthetic polymers (PA, PE, PP, PET PTFE) with treatment in $O_2$-, air-, $NH_3$-plasma. Hydrophilic treatment serves also as dirt-repellent and antistatic finish. Hydrophobic finishing of cotton, cotton/PET, with treatment with siloxan- or perfluorocarbon- plasma. Oleophobic finish for cotton/polyester, by means of grafting of perfluoroacrylat.

*4) Dyeing and printing.* Improvement of capillarity in wool and cotton, with treatment in oxygen-plasma. Improved dyeing polyester with $SiCl_4$-plasma and for polyamide with Ar-plasma.

*5) Other properties.* Bleaching wool, treatment oxygen-plasma. UV-protection dyed cotton/polyester, with treatment HMDSO in plasma. Flame-retardant feature for PAN, Rayon, cotton, treatment: e.g. phosphorus containing monomers.

*6) Metal-Coated Organic Polymers.* Metal-coated organic polymers are used for a variety of applications. If the metallised polymer is expected to fulfil its function, it is essential that metal strongly adheres to the polymer substrate. This can be obtained with a plasma pre-treatment of the polymer.

*7) Composites and Laminates.* Good adhesion between layers in laminates depends upon the surface characteristics of fibres in layers and the interactions taking place at the interface. A prerequisite condition of good adhesion remains the surface energy of fibres, which can be modified with plasma treatments.

*8) Applications in Biology and Medicine.* Fabric favouring overgrowth with cells for cell culture tests, fermentation or implants. Fabric not favouring overgrowth with cells for catheters, membranes, enzyme immobilisation, sterilisation.

*9) Applications in Membrane and Environmental Technology.* Gas separation to obtain oxygen enrichment. Solution-Diffusion Membranes to obtain alcohol enrichment. Ultra filtration membranes to improve selectivity. Functionalized membranes such as affinity membranes, charged membranes, bipolar membranes.

Let us start with a more specific discussion of some cases studies. First of all we will study the role of plasma treatment on natural fibres such as wool and cotton. Then Nylon 6 is discussed and a brief review of plasma treatment on other synthetic polymeric fabrics is performed.

## Plasma treatment of wool

There is an enormous potential in the plasma treatment of natural fibre fabrics. Plasma treatment has proved to be successful in the shrink-resist treatment of wool with a simultaneously positive effect on the dyeing and printing.

The morphology of wool is highly complex, not only in the fibre stem but also on the surface as well. It is in fact the surface morphology to play an important role in the wool processing. Unwanted effects such as shrinkage, felting and barrier of diffusion are most probably due to the presence of wool scales on the fibre surface. In the past, the modification of wool surface morphology were conducted either by chemical degradation of scale (oxidative treatment using chlorination) or by deposition of polymers on the scale [15,16]. However, in both processes, a large amount of chemicals generated from incomplete reactions polluted the effluent. The oxidation is also required to reduce the hydrorepellance of wool to obtain good dyability.

Wool is composed at 95% of a natural polymer, the keratin. In the outer part, the cuticle, the cells are in the form of scale (see a drawing in Fig. 4).

Cuticle cells overlap to create a directional frictional coefficient: the scales are moved by water and they have the tendency to close and join together with the typical movement that is proper to have a good textile but it is also producing felting and shrinkage.

Plasma treatment of wool has a two-fold effect on the surface. First, the hydrophobic lipid layer on the surface is oxidised and partially removed. Since the exocuticle, that is the layer of the surface itself (epicuticle), is highly cross-linked via disulfide bridges, plasma treatment has a strong effect on oxidising the disulfide bonds and reducing the cross-link density.

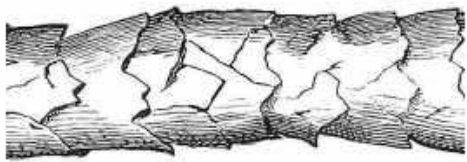

Fig.4: A wool fibre in covered by cuticle scales. Very long cells are building the inner structure of the fibre.

The former plasma treatments on wool were done with the corona discharge but it was not giving a uniform treatment on the fabric: the cuticle is modified, being formed on the fibre microroughness and holes. The corona discharge, consisting of a series of small lightning-type discharges, has the advantage to be easily formed at atmospheric pressure by applying a low frequency high voltage over an electrode pair. Corona discharge is usually inhomogeneous and then problematic for textiles. Rakowsky compared corona and glow discharge plasma concluding that the second is better: both treatments are involving only a surface thickness ($\sim 10^{-8}$ m) and then do not modify the wool structure [17].

As the surface is oxidised, the hydrophobic character is changed to become increasingly hydrophilic. The chemical and physical surface modification results in decreased shrinkage behaviour of wool top; the felting density decreases from more than 0.2 grams per cubic centimetre to less than 0.1 grams per cubic centimetre. After plasma treatment the fibre is more hydrophilic, then a layer of water can be formed during washing procedures with a reduction of friction among fibres and a consequent felt reduction.

With respect to shrink-resistant treatment, this effect is too small compared with the state-of-the-art chlorine/Hercosett treatment. Therefore additional resin coverage of the fibre surface is required. It should be mentioned that the plasma treatment brings additional advantages, in particular increasing dyeing kinetics, an enhanced depth of shade, and improved bath exhaustion.

Rakowsky discussed treatments with plasma gases of $O_2$, air, $N_2$ at low pressure: he observed a regular abrasion of the surface, the removal of the fat acid layer, the reduction of aliphatic carbon (C-C, C-H) of about 20-30% and the appearing of carboxylic COOH groups [17,18]. With $N_2$ plasma the better effect on the dyeing of wool is obtained: in fact, it is producing amine groups on the surface possessing dye affinity.

To obtain measurements of the role of plasma on the dyeing processes, the coloration of a dye bath is evaluated in order to obtain the so called "exhaustion curves", that give the kinetic behaviour of the dyes during a dyeing process as function of the dyeing process parameters. Figure 5 shows the exhaustion curves for different plasma treatments on wool [19].

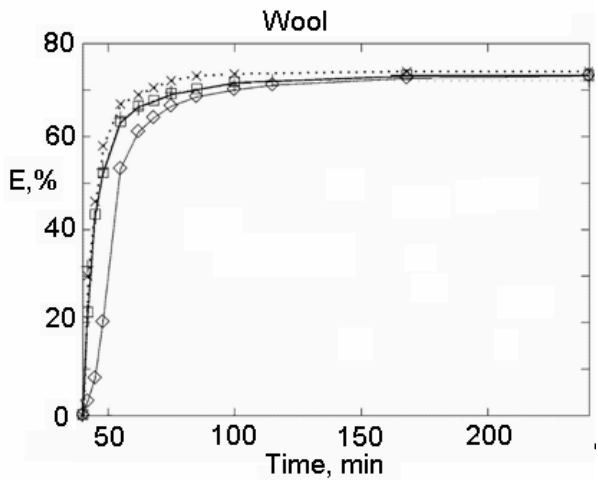

Fig.5: Exhaustion curves for different plasma treatments, as a function of treatment time [19]: diamonds are showing data for control sample without plasma treatment, + for $O_2$ plasma, × for $N_2$ and squares for the mixture $O_2/N_2$.

**Hydrophobic finish on cotton**
Cotton is a major crop in many countries and it is still the most important textile fibre in the world despite market inroads made by synthetic fibres such as polyester. Fabrics from cotton are comfortable to wear and can be dyed a wide range of attractive colours. Each cotton fibre is an elongated cell: properly it is a seed hair, which grows from the seed in a closed seedpod called a boll. When the boll opens, these tubular fibres are exposed to air, lose moisture, and collapse to a flattened, twisted structure. The mature cotton fibre is actually a dead, hollow cell wall composed almost entirely of cellulose. The lengths of single cotton fibers vary, generally about one inch. It is important to understand the relationship between the structure of this unique natural fibre and its properties. Many of its features are too small to be seen with optical microscopes. Electron microscope gives beautiful images of cotton structure (see Fig. 6). The scanning electron micrograph can shows the extreme difference between length and width of fibre and the flattened, twisted shapes formed when the fibre dried.
A variation in the structures of fibre cross sections is present. Since cotton fibres are natural products and then quite different from each other: when they grow, their hollow tubes fill with cellulose and when cotton is harvested, some fibres contain more cellulose than others. Fibres with nearly full tubes have somewhat bean-shaped cross-sections, but fibres with tubes that are not filled with cellulose are flatter.

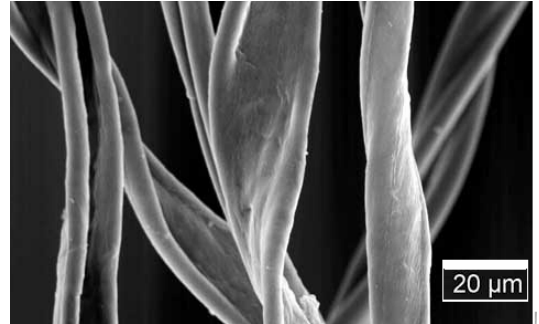

Fig. 6: Cotton flat fibres.

The outer surface of the fibre, known as the cuticle, contains fats, waxes and pectins that confer some adhesive properties to the fibres, which together with natural twist of the fibre means that cotton fibres are well suited to spin into yarns. When cotton yarns and fabrics are desized, scoured and bleached, the cuticle is removed and it is possible to obtain fibres with very high cellulose content. In fact, cotton fibre is mostly cellulose, which is a polymer of many individual glucose molecules. As we have previously told, a modification of the mechanical properties of the cotton fibre can be obtained by oxygen plasma. As in the case of wool, plasma modifies the surface of the cotton fibre, adding wettability.
Let us discuss how to induce a hydrophobic feature on the textile cotton surface. The textile surface has then the property to be less sensitive to spots. In this case, we need two processes: activation and polymerisation. It is depending on the power of source energy and pressure, on the time of exposure and on the plasma gas under use.
The research group of M. McCord at the North Caroline State University used $CF_4$ and $C_3F_6$ on cotton denim fabrics with a low-pressure low-temperature plasma system to increase hydrophobic properties of the surface [20]. To determine the hydrophobicity, contact angle and wet-out time measurements were conducted and the effectiveness of $CF_4$ and $C_3F_6$ gases compared using atomic chemical composition as well as XPS analysis. Plasma treatments were performed on cotton denim fabrics in a RF (13.56 MHz) plasma chamber in a capacitively-coupled mode with the substrates placed on the lower electrode. During the treatments, pressure, RF input power, and exposure time were varied according to an experimental test matrix. After the plasma treatment, the researchers evaluated the surface wettability by means of the sessile drop technique, where a distilled water droplet is placed on the fabric surface and observed through a telescope and the contact angle of the droplet on the surface of the fabric was measured. If we observed a greater contact angle, the greater is the surface hydrophobicity. Contact angles of denim fabrics treated in $C_3F_6$ plasma increased with power and exposure time but decreased with increasing pressure. Overall, the hydrophobicity of treated desized denim fabrics was higher than that of treated sized denim fabrics. This indicates that sizes on the denim fabrics play a role in determining surface wettability even after fluorination in a $CF_4$, $C_3F_6$ plasma treatment.

After $CF_4$ and $C_3F_6$ plasma treatment, $-CF_x$ hydrophobic chemical groups were obtained by the chemical reaction between surface molecules and fluorinated gases. The other carbon chain groups (hydrophilic groups) decreased, and their composition rates were changed according to plasma conditions. Measurements on the water contact angle for plasma treatment with $CF_4$ on cotton, PET and silk are given in Ref. [21]: for cotton the angle changes from 30 degrees to angles ranging from 90 to 150 degrees, for different pressures and times of plasma treatment. On PET, the angle goes from 105 degrees to 120-155 degrees. $CF_4$ plasma gives to the PET surface a structure similar to Teflon, then with a very high water repellence [22]; instead, a treatment with $O_2$ gives to PET an hydrophilic surface. Examples of contact angles are shown in Fig.7: wettability is obtained for angles less than 90 degrees. Let us remember that a liquid is wetting a surface if the surface energy of substrate is greater than that of liquid.

**Nylon 6 plasma processing**.
Polyamides, also known as nylon, are the most widely used semi-crystalline engineering thermoplastics and also yield excellent fibre [23]. Nylons are characterized by their good thermal stability, flexibility and mechanical properties and are the most important synthetic textile fibers originally used for women's stockings. Nylon 6 does not behave much differently from nylon 6,6: the only reason both are made is because DuPont patented nylon 6,6, so other companies had to invent nylon 6 in order to get in on business. The monomer unit is coming from caprolactam.

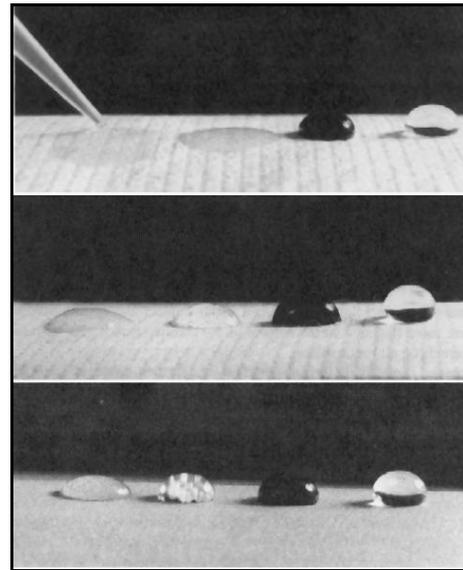

Fig.7: Four liquids (from left to right: isopropyl alcohol, mineral oil, synthetic blood and water) on surfaces of three materials (from top to bottom, woodpulp/polyester, polyester, polypropilene): according to the surface energies, liquid drops are showing different contact angles.

The research on the plasma treatment on polyamide is mainly dyeability [24,25], wettability [26] and surface properties [27,28]. Oxygen and air plasma are used to increase wettability and dyeability. On PA6, an air plasma treatment was performed in an industrial production process: an increase of bondability was observed changing the bond from adhesive to cohesive. Moreover, with the same industrial procedure, with an ammonia plasma, a slight increase of wettability was observed [29].

Nitrogen-containing plasmas are in fact widely used to improve wettability, printability, bondability and biocompatibility of polymer surfaces [30,31] and many application of nitrogen containing plasmas for surface modification of different polymers have been investigated. For example, to improve the interfacial strength between polyethylene fibres and epoxy resins, which are cured by amino crosslinking, amino groups were introduced on the fibre surface to promote covalent bonding [32]. Plasma treatment with $NH_3$ give rise to N-functionalities, such as amino (-$NH_2$), imino (-CH=NH), cyano (-C≡N), plus oxygen-containing groups due to post plasma atmospheric oxidation. Although PA6 adsorbs more water than other frequently used synthetic polymers, it is still classified as hydrophobic in many applications. When water is used as testing liquid, it can be observed that the introduction of polar groups on the surface due to $NH_3$ plasma treatment makes PA6 less hydrophobic.

Nylon 6 fabrics were treated, as discussed in Ref.[33], with low temperature plasma with three non-polymerizing gases (oxygen, argon and tetrafluoromethane) to investigate mechanical and thermal properties. After plasma treatment, the properties of the fabric, including surface morphology, low-stress mechanical properties, air permeability and thermal properties, were measured by researches, observing that nylon fabrics treated with different plasma gases exhibited different morphological changes. Low-stress mechanical properties revealed that the surface friction, tensile, shearing, bending and compression properties altered after the treatments. The changes in these properties are believed to be related closely to the inter-fibre inter-yarn frictional force induced by the plasma treatment. It was also observed a slightly decrease in the air permeability of the treated fabrics, probably due to plasma changing the fabric surface morphology. A change in the thermal properties is in agreement with the above findings and can be attributed to the amount of air trapped between the yarns.

For what concerns $CF_4$, it is a non-polymerizing gas that does not polymerize itself, but tends to form thin films on the fibre surface subjected to the glow discharge. Gazicki et al. [26] examined $CF_4$ plasma –treated fabrics and found that ablation was accompanied by the deposition of thin films on the fibre surface. Yip et al. [33] suggested that a shorted exposure time will favour polymerization while a longer exposure time will favour ablation. In Ref. [33], the authors present images where it is possible to see small deposits on a fiber for short time treatment with $CF_4$.

**Activation of PP, PE, PET and PTFE.**
Polypropylene (PP) is a very interesting material for plasma treatment: it is a very hydrophobic material with extreme low surface tension. On the other hand, PP is used in a large number of technical applications where an improved wettability or adhesion properties are advantageous. This is also the case of PP technical textile applications such as filters for medical applications. Since PP non-woven filters can be wetted only with liquids with surface tension <35mN/m, no water can pass through the PP-web without applying a high pressure: using an oxidative plasma with a short treatment time can greatly improve wettability.

Varying treatment time, vacuum level and treatment power of a $O_2$ plasma, it was observed [34] that the increase in surface tension of PP is not in correlation with the intensity of plasma treatment. An increase in wettability can be indeed observed but only at relatively low treatment intensities: once the optimum is reached a sharp drop in wettability was obtained if the plasma treatment intensity is raised further.

Improvement of wetting in PP has been observed for air and $NH_3$-plasma. As in PP, also in polyethylene PE, polyethylene terephthalate PET and polytetrafluoroethylene PTFE, treatments in air-, $O_2$- and $NH_3$-plasma are usually performed. These treatments are able to increase wettability but also adhesion for these materials (hydrophobic finishing on cotton/PET and PET already mentioned, are produced with siloxan- or perfluorocarbon-plasma [35]). In the Table I, data on the modification of surface energy and contact angle of water on several polymeric substrates are shown.

But let us talk about another very important problem, that is how to increase adhesion of different polymer-metal systems, namely, PET-Al, Kapton-Al, and Teflon-Cu. To this purpose both reactive ($O_2$ and $NH_3$) and inert (He) gases have been used for the plasma treatments. The results obtained with PET are particularly interesting from both the fundamental and the industrial points of view. $NH_3$ plasma treatments show to be successful in obtaining higher PET-Al adhesion values at very short plasma duration: the shorter the $NH_3$ treatment time the higher is the adhesion increase. A treatment of 0.1 s is sufficient to promote a 15-20 fold increase of adhesion [35].

Table I

Surface energy and water contact angle before and after plasma treatment

|  | Surface Energy Dynes/cm | | Water contact Degrees | Angle |
| --- | --- | --- | --- | --- |
|  | Before | After | Before | After |
| Polypropylene PP | 29 | >73 | 87 | 22 |
| Polyethylene PE | 31 | >73 | 87 | 42 |
| ABS | 35 | >73 | 82 | 26 |
| Polyamide PA | <36 | >73 | 63 | 17 |
| Polyester PET | 41 | >73 | 71 | 18 |
| PTFE | 37 | >73 | 92 | 53 |

The driving force of metal/polymer adhesion is the acid/base character of the interaction, promoted by introducing basic functionalities onto the PET surface. In fact, the selective basic strength of N-containing groups, is very well correlated to the polymer/metal adhesion, provided the treatment time is shorter than 30 s. With a longer treatment time, a weak boundary layer forms on the PET surface. The highest adhesion is obtained at 0.1 sec. This means that $NH_3$ plasma treatments are suitable also for web industrial coating processes.

**Industrial devices**
J. Reece Roth describes in his book [36] the industrial processes that use plasmas or plasma related technologies. He talks of plasma sources, plasma reactor technologies and applications of thermal and non thermal processes. Plasma are industrial useful because they possess at least one of two important characteristics. The first is the high power or energy density (in RF inductive plasma torches, the plasma energy density is ranging from 100 W/cm$^3$ to above 10kW/cm$^3$) in the case of thermal equilibrium plasma. The industrial applications are vaporizing bulk materials, welding, flame spraying, and high temperature processing.
The second major characteristic of plasma relies upon the production of active species, that are more numerous, different in kind and more energetic than those produced in chemical reaction. As we have seen, the non-thermal plasma, obtained in the glow discharge, produces active species by which it is possible to do things on the surface of materials that con be done in no other way. A detailed description of industrial devices can be find in [36]. In this section, a list of societies, which commercialise low-pressure plasma devices is given. In U.S.A. the 4$^{th}$ State Inc., Belmont California [37], rents and sales plasma devices at 13.56 MHz (the publications on the web site of 4$^{th}$ State give good detailed description of plasma processes, see for instance [38]), but the company is not producing devices for textile treatment.
Europlasma [39] is a Belgian manufacturer of low pressure gas plasma systems: it is producing the so called Roll-to-Roll textile treaters and designed batch reactors according to client requests. CD Roll 1800 is a Roll-to-Roll plasma treater used for the treatment of non woven and web material to activate the surface prior to lamination, to improve wettability and adhesion, and for a hydrophobic/oleophobic finishing with a plasma polymerisation. The device works at 40KHz.
In Italy, HTP Unitex produces and sales, but also uses for surface activation in industrial productions, roll-to-roll devices working at 10KHz [40]. As active plasma species they are using $O_2$, $N_2$, $NH_3$ to increase wettability and bonding ability to produce sportwear fabrics. As it can be seen in Fig. 8, the fabric is driven by a roll system, to pass among a set of rods that are electrodes generating plasma. Rolls and rods are inserted into the vacuum chamber, and special control devices must be used to monitor stresses in fabrics. Looking in the chamber, during the plasma treatment, it is possible to see the glow discharge among electrode rods (see fig.9).

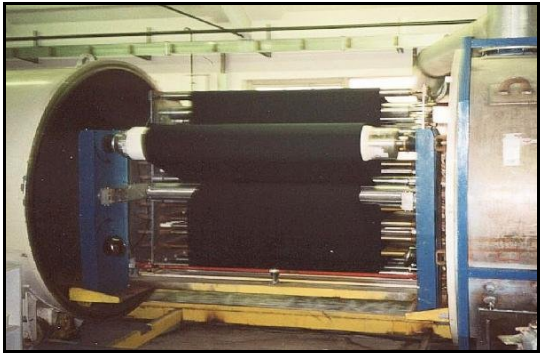
Fig. 8 HTP-Unitex roll-to-roll system.

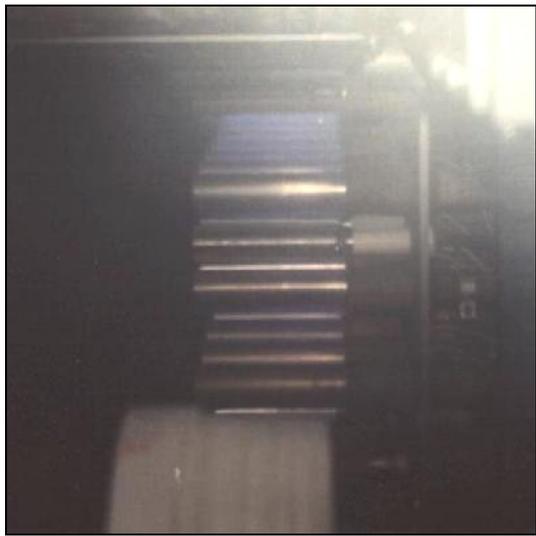
Fig.9 Glow discharge in the vacuum chamber of HTP-Unitex

Plasma treatments, largely used in semiconductor, mechanics or packaging industry, are now starting to receive increasing consideration in the textile industry, since the technology is becoming more convenient when compared with procedures harmful to environment. It is just a matter to acquaint plasma with textile industry. Plasma is relatively new for the textile industrial sector, traditionally driven by chemical processing. Low-pressure plasma systems, the oldest and more studied systems, can be used to induce hydrophilic and hydrophobic effects on textiles. Unfortunately they are requiring expensive vacuum equipments. Plasma is then not seen as a real advantage. The textile industry is actively searching for plasma systems, which operate at atmospheric pressure. Atmospheric plasma systems are well known and currently used in industries devoted to processing packaging materials. Several industrial devices, from standard corona to afterglow discharge systems, are then ready for use on textile surfaces.

**Glow discharge plasma**
Let us consider partially ionised gas. When a sufficient high potential difference is applied between two electrodes placed in the gas, a breakdown among electrons and positive ions produces a discharge. Excitation collisions, followed by de-excitation, give the characteristic luminescence that is seen as the glow of the discharge. Due to collision processes, a large number of different plasma species are then generated: electrons, atoms, molecules, several kinds of radicals, several kinds of (positive and negative) ions, excited species, etc. Different species are in interaction with each other, making the glow discharge plasma a complicated gas mixture [41]. When materials are subject to plasma treatments, the subsequent and significant reactions are based on free radical chemistry. The glow discharge plasma is efficient at creating a high density of free radicals by dissociating molecules through electron collision and photochemical processes. These gas-phase radicals have sufficient energy to disrupt chemical bonds in polymer surfaces on exposure, which results in formation of new chemical species.

To sustain a glow discharge in direct current, electrodes must be conductive. In the simplest case, a discharge is formed by applying a potential difference (from 100 V to few kV) between electrodes inserted in a cell filled with a gas at a pressure ranging from mTorr to atmospheric pressure. When one or both electrodes are non-conductive, then an alternating voltage is applied to each electrode, usually in the range of radio frequency. Concerning gas pressure, the glow discharge can operate in a wide range of these pressures. In the case of a low- pressure operating gas, the volume of the discharge can be rather expanded. Operating at atmospheric pressure (atmospheric pressure glow discharge, APGDs), the linear dimension over which the discharge develops is reduced. The typical dimension is 1-2 mm, reducing the characteristic length of the discharge chamber. Stable APGDs used for technological applications have been developed, differing with respect to the structure of electrodes, carrier gas and operating frequency. Typically, APGD systems are characterized having one electrode covered with a dielectric with the discharge operating in alternating voltage.

The type of discharge gas determines the stability of the glow discharge. Helium gives rise to a stable and homogenous glow discharge, whereas nitrogen, oxygen and argon require higher voltage for ionization and can cause the transition to a filamentary glow discharge. However, it is still possible with a change in the electrode geometry to obtain a somewhat homogenous, yet somewhat filamentary, glow discharge [42].

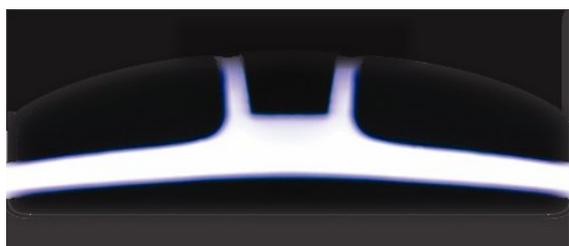

Fig.10 Homogeneous atmospheric glow discharge between electrodes in Plasma$^3$ [43]. Films or fabrics are running during the treatment over the lower electrode.

The requirement of vacuum systems for developing low-pressure plasmas has been a burden for textile industry, hence plasma generated at atmospheric pressure is more appropriate. Two main classes of atmospheric pressure plasmas are, according to Reece Roth, dielectric-barrier discharge (DBD) and atmospheric plasma glow discharge (APGD). Sometimes DBD are also known as "corona treaters": DBD devices are well known in the packaging industry, where they are used to increase the wettability of polymeric films. They are strongly related to the APGDs, which operate with an a.c. voltage of 1-100 kV at a frequency of few Hz to MHz. APGDs are devices having homogenous and uniform discharges across the electrodes, whereas the DBDs produces discharges with micro-discharge filaments and considerably less uniform. In Fig.10, the glow discharge in the APGD system (Plasma$^3$), developed by the Enercon Corporation, is shown [43].

**From vacuum to atmospheric pressure.**
Russian researchers were the first to develop a full, industrial scale roll-to-roll vacuum reactor with plasma in the glow-discharge family. They succeeded to reach large scale in fabric width plasma reactors, applying in the building of the vacuum chambers technologies developed for space missions. Such reactors can treat any type of fabric and are only limited in the amount of material that can be treated in one batch (i.e., the fabric roll diameter must be limited, due to the chamber dimension and to mechanical problems with roll-to-roll systems). The main drawback of the system is the need for a vacuum chamber, a rather expensive part of the equipment.

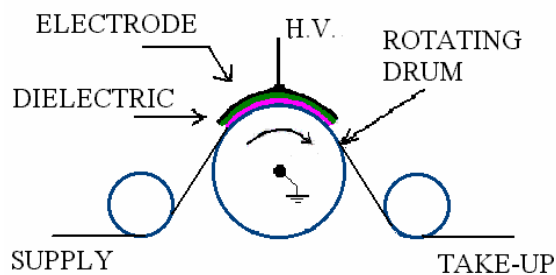

Fig.11 Taken from a supply reel, the film undergoes a DBD treatment between a grounded rotating drum and a high voltage electrode. The other is covered with dielectric.

Continuous atmospheric plasma treaters are currently being used in the printing processes of packaging films where the surface treatment is achieved with a film velocity reaching 100 m/min. or even greater. In textile industries, such high velocities are not required, then an atmospheric plasma treater mounted directly on the loom is not necessary. It can be easily integrated into textile finishing lines. For this reason, vacuum technology is regarded as being non-competitive. Extra requirements such as the ability to treat full width textiles of at least 2 m wide, and higher processing speeds at 20 m/min are more economically addressed by atmospheric plasma treatment systems. In Fig.11, a schematic diagram for film treatments with a DBD system is shown, in the usual set-up to increase wettability. From a supply reel, the film passes through the glow discharge, driven by the rotating drum.

Many DBD systems are currently used for packaging production and new APGD devices are available. The driving force in the packaging industry for development of atmospheric plasma sources is the unavoidable requirement of a continuous, roll-to-roll technology. Production processes in the textile industry are essentially non-continuous and so this drag force is lacking. Let us explore the advantages of atmospheric plasma technology for the treatment of textiles.

**Atmospheric plasma treatments for textiles.**

First of all, let us remember that the atmospheric plasma surface treatment with corona treaters is used to improve wettability of polymeric films. As we have seen, the increase of wettability is one of the first and well-known effect obtained on textiles with vacuum reactors with $O_2$-, air-, $NH_3$- plasma, together with an enhancement of capillarity of wool and cotton fabrics. In fact, the character of the wool surface can be changed from hydrophobic to hydrophilic, obtaining a reduced felting of tops and fabrics by means of simple corona treaters. Another effect that can be easily obtained with a corona treatment is an overall cleaning of surfaces, due to the ion bombardment in the discharge. Hydrophobic properties, as observed on cotton fabrics can be also induced with a proper choose in type and proportion of gas chemistries used in the reactor [44]. When the processing gas is air, the processing costs immediately favour atmospheric reactors. In the case where a pure processing gas, for instance ammonium or hydrogen, is required, processing is obtained at the lowest effort in vacuum reactors. But today, supply of processing gas and evacuation of waste gas has been engineered to be very economical requires in atmospheric pressure reactors.

Let us consider cotton again, and in particular, the objective of increasing hydrophilic features and reducing chemical waste of existing pre-treatment processes for cotton fabrics. Researchers in Portugal used a corona discharge in an air atmosphere and showed that corona is a very effective way of increasing hydrophilicity without affecting the integrity of the fibre or yarn [45]. The treatment leads to chemical and physical changes to the waxy cuticle layer of the cotton without damaging the cellulose backbone. Increased wetting of fibers was observed, together with a decrease in the pH of water in contact with the cotton.

With corona discharge, a similar effect over the waxy cuticle is expected as that obtained with softening finishing. If a corona discharge is made before the application of the softening agent, more favourable conditions for fixation are created, either increasing reaction with the fibre or increasing physical conditions for penetration. Softener content in effluents will be lower.

The increase of hydrophilicity of cotton is important from another point of view. Cotton warp yarns have to be sized prior to weaving to apply a protective coating to improve yarn strength and reduce yarn hairiness. Starch-based products are most frequently used for slashing cotton yarns. Carboxymethyl cellulose (CMC) and polyvinyl alcohol (PVA) are the next most frequently used sizing agents for cotton yarns. PVA is used primarily for slashing synthetic yarns, and is also often used as a secondary sizing agent to starch for cotton yarns. Size materials must be removed by desizing prior to dyeing and finishing woven fabrics. Sized fabrics must also be subjected to washing with water at temperatures around 90°C in order to effectively remove size. A complete PVA size removal is difficult, with many disadvantages such as high energy and water consumption.

In [46] an APGD reactor is used with air/$O_2$/He and air/He plasma. A percent desizing ratio PDR of 99% was obtained with both air/He and air/$O_2$/He plasma treatments followed by cold and hot washing: as a conclusion, atmospheric plasma treatment may greatly increase the solubility of PVA on cotton in cold water. SEM observations revealed that, for both air/He and air/O2/He plasma treated fabrics, the fibre surfaces were nearly as clean as the unsized fabric, indicating that almost all PVA on cotton was removed. This agrees with the PDR results. In contrast, there was a substantial amount of PVA remaining on the fibre surfaces when fabrics were only cold and hot washed, while much less PVA was left on the fabric desized by $H_2O_2$.

We already discussed wool and its complex scale structure on the fibre surface. Atmospheric plasma, both corona and APGD methods [47,48], are suitable for wool too. As for cotton, if a corona discharge is made before the application finishing antifelt agent, more favourable conditions for agent fixation are created and chemical compound content in effluents is lower.

For all natural and synthetic fibers, the penetration of activating species in APGD plasma into materials is so shallow that the interior of the material is only lightly affected. Plasma treatment can be used for modifying the surface properties of several synthetic fibres. Among these fibres, polypropylene-based fabrics have an inherently high crystallinity and are the most challenging of fabrics with which to promote dyne uptake, ink and coating adhesion. In [49,50], an APGD reactor was used to promote the adhesion of water-based inks to a .40 mil. polypropylene fabric in a roll width of 1524mm. The initial surface tension level was measured at 31 dynes/cm. The surface tension was increased to 52 dynes/cm using an APGD reactor with an air/$O_2$/He plasma, and with a corona treater. The corona and APGD materials were then printed via flexography with four colours (non-processed) using an anilox roll of 275 lines/cm with a 2.1 cell volume. The ink was transferred to the fabrics at 60 mpm and then forced air dried in-line at 45º C. A 12mm x 50mm tape (ASTM specified) was adhered to the printed surfaces in multiple locations on untreated, corona treated and APGD treated surfaces and allowed to dwell for 60 seconds. Following the dwell period, the tape was removed and revealed 0% ink surface retention on the untreated material, 90% ink surface retention on the corona treated material, and 100% ink surface retention on the APGD treated surface.

A deep investigation on several polymeric fabrics with DBD treatments is reported in [51], for a discharge of random filamentary type, the more convenient and economic alternative for the controlled modification dependent on surface oxidation. Researchers characterized the dielectric barrier discharge in terms of time, inter-electrode gap and energy deposited at the processing electrodes, as process variables. The treated samples were investigated using contact angle/wickability measurement and scanning electron microscopy. Of the three process variables investigated, the duration of the treatment was found to have a more significant effect on the surface modifications than that obtained varying the discharge energy or the inter-electrode gap.

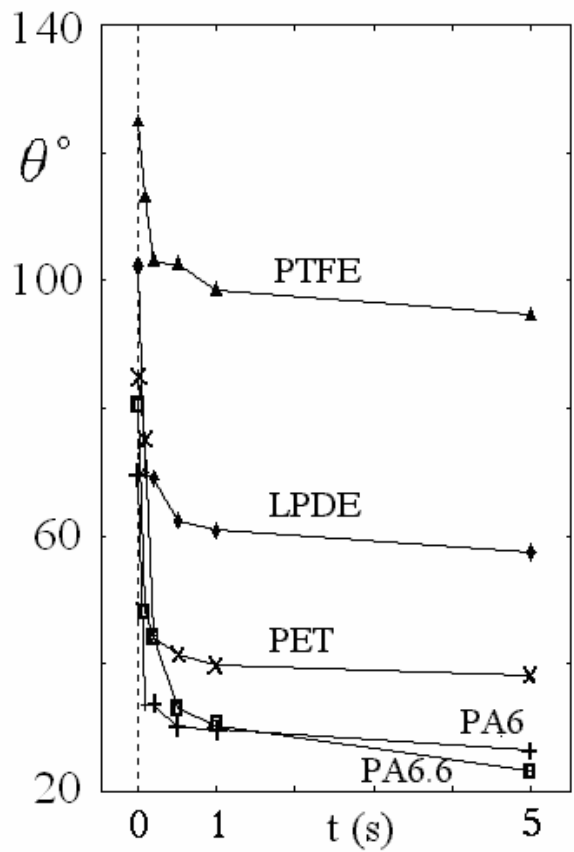

Fig.12 Contact angles of water on several materials as a function of the treatment time in a DBD plasma device, from [51].

Very short air-DBD treatments (fractions of a second in duration) markedly and uniformly modified the surface characteristics for all the materials treated, to the effect that wettability, wickability and the level of oxidation of the surface appear to be increased strongly within the first 0.1–0.2 s of treatment. Any subsequent surface modification following longer treatment (>1.0 s) was less important. The modification of the surface properties also appears to be stable with time. In Fig. 4 the behaviour of the contact angle of water for several polymeric materials is shown. And in the following fig.12, some drops of liquids on different surfaces are shown to see the contact angle changes according to the nature of liquids and surfaces.

In [49], a very interesting SEM analysis of polyester fabrics gave complementary valuable information on the role of the additional polymer film placed on the grounded electrode during surface processing. Images show that if samples are placed during treatment directly on the grounded aluminium cylinder a strong, but very localized, degradation of the fabric takes place, with fibres melting at the locations were the bundles of fibres are crossing each other. The effect is most probable due to the nonhomogeneity of the discharge, with the micro-discharges tending to localize on spots where they can reach directly the grounded electrode directly. This behaviour can be seen as a type of corona effect where the channels at the yarn crossover points in the fabric promote a localized more energetic discharge than the norm. Results are very different if the fabric samples are placed on the additional thin polymer (polyethylene) film, so the grounded electrode is electrically isolated from the discharge. In this case, no visible degradation of the fabric takes place, with no visual difference compared to the untreated case, even for extended treatment duration.

**Conclusions.**
Let us conclude telling the extra advantages of plasma treatments. The finished textile shows better performance and improved colour fastness properties. Though currently not very relevant in produced amounts, this type of high-performance textile will certainly grow in economic importance. As a result of their high added value even small textile batches can be produced at high profit, although perfect process control is absolutely necessary. Typically, textiles for medical applications or uses in the sector of biotechnology are expected to increase in importance. Key future applications such as special selective filtrations, biocompatibility, and growing of biological tissues, would be interesting fields for plasma physics.

**Acknowledgment**
Many thanks to Ermanno Rondi (President U.I. Biella) for useful discussions.

*For a very short history of plasma, see A. Sparavigna, Physics in Carnacki's investigations, arXiv:0711.4606v3 [physics.pop-ph]*